# Building Bridges: Julia as an MLIR Frontend

Jules Merckx

Supervisor: Prof. dr. ir. Bjorn De Sutter
Counsellors: Dr. ir. Tim Besard, ir. Thomas Faingnaert

*Abstract*—Driven by increasing compute requirements for deep learning models, compiler developers have been looking for ways to target specialised hardware and heterogeneous systems more efficiently. The MLIR project has the goal to offer infrastructure that can be used to develop new compilers and represent code at different levels of abstractions. While MLIR excels at offering developers a way to write new intermediate representations (IRs) and transformations, there is no easy way for end users to generate code in these IRs. In this work, we explore using the Julia programming language as a high-level input language for generating MLIR code. Most importantly, we focus on extensibility, allowing package developers to implement bindings to MLIR dialects in an intuitive and easy-to-use manner. By building on the Julia programming language, and its expressive features such as multiple dispatch and its extensible compiler, we design and implement a framework to generate MLIR code. Additionally, we evaluate this framework in three case studies. Ranging from developing a small domain specific language (DSL) for einsum expressions, to specifying transformations on MLIR code and programming kernels to be run on graphics processing units (GPUs).

## I. Introduction

Among other factors, the explosive success of deep learning in recent years has led the field of computer science to shift more towards the use of heterogeneous or specialised hardware [1]. At the same time, programmers expect to keep working at the same high levels of abstraction to develop and execute new deep learning models, abstracting over underlying hardware details and enabling code portability, while still reaping the benefits of new hardware and architectures. The schism between high-level programming and efficient hardware utilisation obviates the need for new and more powerful compiler techniques, and many projects have appeared, trying to improve the status quo.

Developing new compiler optimisations, however, requires not only extensive knowledge of the domain of the compute workload, but also of the intricate details of the particular compiler being improved. Take for example LLVM [2], a compiler framework used by compilers for many programming languages such as C, C++, Julia, Swift, and Rust. Its codebase consists of millions of lines of code written in a variety of languages. A typical end user lacks the expertise to implement new features or fixes they need, and is reliant on the goodwill of expert compiler developers, who may have differing priorities or simply lack the time to implement those features. To add insult to injury, different projects, be it programming languages, deep learning compilers, or other tools requiring compiler technology, often re-implement much of the same functionality such as tooling for static single-assignment (SSA) manipulation, constant propagation, dead code elimination (DCE), and other fundamental compiler passes. This duplication of effort leads to less time being available to develop compilation methods specific to the target domain, potentially leaving a lot of performance on the table.

One of the projects trying to tackle this problem is MLIR [3], which aims to provide a generic framework to build and interoperate between compilers. It offers tools to create new IRs and has infrastructure in place to specify optimisations and transformations between IRs efficiently. Projects exist to generate MLIR code from a host of different sources, from DSLs such as Triton [4], to full-blown programming languages such as Mojo [5]. Despite differences in approach, most of the existing projects use Python, or at least syntax heavily inspired by it, as a source language. This work explores using the Julia programming language as an alternative. Julia is a strong contender for compiler-related research because of its extensible compiler, allowing programmers to hook into and change the behaviour of many parts of the compiler that are typically inaccessible in other languages. In this work, we will discuss the design and implementation of a framework that helps developers to create high-level interfaces in Julia for users to generate specialised MLIR code.

## II. Background and Related Work

### A. MLIR

MLIR, or multi-level intermediate representation, is a general framework for compiler development. It was initially developed to solve the problem in the machine learning field, where modern frameworks, such as Pytorch [6] and TensorFlow [7], all implement their own compilers to transform and optimize programs. This leads to a lot of redundant work by framework developers and moreover, leads to friction when attempting to interoperate between frameworks [3]. The vision of MLIR, which is part of the LLVM project, is much broader, though. As illustrated in Figure 1, the same observation as for machine learning frameworks can be made for many modern programming languages: they depend on LLVM as an architecture-independent compilation target, yet they all implement their own SSA-based IR. These IRs are used as a bridge between the language-specific abstract syntax tree (AST) format, and LLVM IR. This intermediary step is necessary to allow for high-level optimisations that are hard to perform on LLVM IR because LLVM IR typically does not suffice to represent all language semantics at a high enough level.

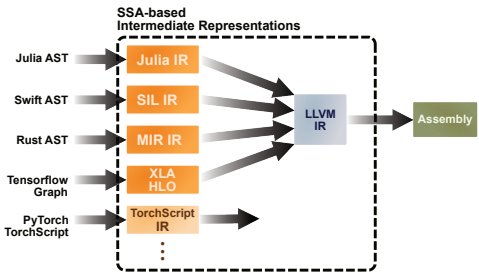

Fig. 1: Many different languages implement their own SSA-based IR as an intermediary target when compiling to LLVM. Note, Pytorch's TorchScript also has an SSA-based representation but does not target LLVM. Figure adapted from [8].

The solution proposed by MLIR is based around *dialects*. A dialect consists of operations, attributes, and types. By defining new dialects, it is possible to extend MLIR with new IR constructs. The official MLIR codebase contains a host of carefully curated dialects covering a wide range of functionality. For example, the `arith` dialect contains basic mathematical operations on integer and floating point arguments, the `linalg` dialect defines linear algebra operations such as matrix multiplications and convolutions on tensor-like types, and the `llvm` dialect is a mapping onto all different types and operations of LLVM IR, essentially providing a bridge between MLIR and LLVM.

Related to dialects are transformations. These can be used to implement domain-specific optimisations on a dialect, or conversions to different dialects. These conversions, especially, are of great importance in MLIR, as they enable the gradual lowering of MLIR code towards machine code.

## B. Julia

The Julia programming language, founded in 2012, is a "Fast Dynamic Language for Technical Computing" [9]. Among other applications, it is being used for deep learning [10], climate modeling [11], mathematical optimization [12], GPU programming [13], and solving differential equations [14]. Because Julia compiles and heavily optimises functions using LLVM whenever type inference is able to fully infer all variable types, it is possible to write seemingly dynamically typed programs while still getting performance rivaling that of ahead-of-time (AOT) compiled languages such as C. [15]

One of the defining features of the Julia programming language is multiple dispatch. As opposed to single dispatch, where the type of a single argument of a function (often called `self`, or `this`, in object-oriented languages) is used to determine which specific method should be called, multiple dispatch decides the method by finding the best match in the method table based on *all* arguments. This conceptually simple feature has far-reaching effects on the language and its ecosystem. Idiomatic Julia packages define functions without many type constraints, allowing users, or definitions in other packages, to extend them as they please.

Another feature that makes Julia an attractive choice for this work is its compiler. Thanks to the fact that, for large parts, the Julia compiler is written in Julia itself, it is possible to extend many aspects of the compilation pipeline easily. First of all, macros can be used to implement AST transformations. Farther in the compilation pipeline, Julia's *AbstractInterpreter* framework can be used to specialise particular methods used during compilation to apply optimisations on the IR. This functionality can also be used to change inlining behaviour, run type inference on custom type lattices, or alter code generation. CUDA.jl [13], for example, uses these methods to customise the compiler to generate code that runs on NVIDIA GPUs. Multiple packages also use these methods to implement source-to-source automatic differentiation engines [16], [17] or static code analysis tooling [18]. In this work, we utilise the AbstractInterpreter framework to implement a small code transformation and customise Julia's inlining policy.

## C. Generating MLIR Code

The key use case of MLIR is to implement new dialects that can be used to represent domain-specific constructs. For developers of DSLs especially, this allows for finer-grained control over generated code. Instead of embedding a DSL in a general-purpose programming language in an effort to reuse its compiler infrastructure or runtime, or even writing a whole new compiler from scratch to get the highest possible performance, an MLIR dialect allows interfacing with the MLIR ecosystem, reusing many of the general optimisations and transformations already defined. Compared to an embedded DSL, this has the advantage of providing full control over the compilation pipeline and is more similar to writing a new compiler from scratch, without the burden of having to implement all the general-purpose compiler infrastructure. However easy it is for a developer to create new domain-specific dialects, their usefulness is limited if there is no easy way to generate MLIR code for these dialects. In this work, the problem of generating MLIR code for arbitrary dialects is addressed.

Today, there are different approaches to generating MLIR code in a dialect. Manually writing the MLIR

code can be viable for small amounts of code and quick experimentation. This approach gives the highest degree of control but is rather labour-intensive, akin to writing LLVM IR or assembly instructions by hand. For compiler developers working on new dialects, this is typically the approach taken, though, for example when writing tests for new operations of the dialect. Some support in code editors exists, providing auto-complete and code navigation, still this approach is far from optimal.

Another approach is to mechanistically build IR using MLIR's application programming interface (API). This API allows to incrementally build MLIR code, by creating basic blocks, regions, and operations. MLIR is a C++ project but also offers a minimal C API allowing any programming language with a C foreign function interface (FFI) to interface with it. This C API is used by higher-level languages to offer an interface to MLIR that is easier to use. In Python, the official Python MLIR API lets users more easily loop over IR by using regular Python loops, for example. In Haskell, mlir-hs [19] leverages pattern matching and other Haskell-specific features to simplify IR building. Beaver [20] similarly offers a higher-level builder API in the Elixir programming language. In Julia, MLIR.jl [21] is a package that interfaces with MLIR through its C API. Apart from direct bindings of all the C functions, much of the same functionality is also offered through higher-level abstractions making IR building much more ergonomic. The different MLIR builder APIs are currently most used for building DSLs. They provide a programmatic way to create MLIR code but are unsuited to quickly build a specific piece of code because of the work required to do so. Each basic block needs to be explicitly allocated, control flow instructions need to be carefully inserted and all the bookkeeping required for doing so is left to the programmer.

DSLs, which can be built on top of these APIs, however, do offer a way to easily generate large amounts of specific MLIR code. These DSLs typically generate MLIR code in one or two specific dialects. Triton [4], a DSL embedded in Python to generate code for GPUs, generates MLIR code in the `tt` dialect, for example. Other DSLs, mostly embedded in Python as well, exist for polyhedral code [22], hardware verification [23], quantum computing [24], database query optimisation [25], and many other use cases. While using a DSL is a valid approach to generate specialised MLIR code more easily, the approach suffers in terms of extensibility. The use of MLIR allows reusing much of the compiler backend infrastructure, but all these different projects have implemented a different frontend, essentially running into the same problem that MLIR is trying to solve, at a different level of abstraction. Creating new DSLs, or altering behaviour of existing ones, requires deep knowledge of the frontend internals, and integrating different DSLs with each other is difficult.

A last approach, which is explored in this work, is to generate MLIR code from a general-purpose programming language. This can have the form a full-blown language compiler, such as the Mojo [5] language that uses MLIR directly as its intermediate representation. Mlir-python-extras [26], on the other hand, starts from Python code and uses its AST and bytecode to generate MLIR code. As not all Python language constructs are supported, one could argue that this is not a full-blown language compiler but rather a DSL that happens to correspond to Python syntax.

*D. Goals of this Work*

In this work, an approach to generate MLIR from Julia code is developed. A primary goal for this work is to allow easily extending the IR generation to support generating IR for new dialects. Extensiblity is at the forefront of the Julia programming language. By reusing and building on parts of the Julia compiler, it can be harnessed for generating MLIR code as well. Just as mlir-python-extras is not meant to be a general-purpose Python compiler, this work does not try to propose a different compiler for the Julia programming language.

III. MLIR Frontend Design in Julia

*A. Interfacing with Dialects*

```
@intrinsic Base.:+(a::f32, b::f32) =
        f32(IR.result(Dialects.arith.addf(a, b)))
```

Listing 1: Definition of an intrinsic function that maps Julia's `+` function on the `addf` operation from the MLIR `arith` dialect.

The primary focus of this work is to allow developers to create high-level interfaces to MLIR dialects. These dialects revolve around operations. To bridge the gap from Julia to MLIR operations, we introduce *intrinsic functions*. These functions can be specified by a developer and contain code that builds an MLIR operation. For an end user, intrinsic functions look and behave just like regular functions, but when MLIR is generated, their implementation is run to build the correct operation. Listing 1 shows the definition of an intrinsic function for the `+` method given two arguments of type `f32`. The implementation calls MLIR.jl's builder function for an `arith.addf` operation, the `IR.result` call returns the SSA value of the operation's result, which is finally converted to an `f32` object. The `f32` type is not just any Julia type, but can be converted to an MLIR type, MLIR's homonymous `f32` type in this case, by calling MLIR.jl's `IR.Type` function. This is illustrated in Listing 2. One last thing that needs to be specified before MLIR code can be generated, is what MLIR operations will be generated for control flow in Julia code. Since MLIR code is generated starting from Julia IR, where high-level, structured control flow has been lowered to unstructured `goto` (or `gotoifnot`) instructions, the developer can implement special functions (i.e., `generate_goto`, `generate_gotoifnot`, and `generate_return`) that generate the desired MLIR

```
struct f32 <: AbstractFloat
    value::IR.Value
end

# Mapping from Julia type to MLIR type:
MLIR.IR.Type(::Type{f32}) = …
```

Listing 2: Similar to how intrinsic functions are used to interface with MLIR operations, types can be created that act, for example, like a floating point, but store an MLIR SSA value in reality.

operations. By default, operations from the control flow dialect, `cf`, are generated.

*B. Generating MLIR Code*

```
sigmoid(x) = 1/(1+exp(-x))
generate(sigmoid, (f32, ))
```
```
func.func @sigmoid(%arg0: f32) -> f32 {
    %cst = arith.constant 1.0 : f32
    %0 = arith.negf %arg0 : f32
    %1 = math.exp %0 : f32
    %2 = arith.addf %1, %cst : f32
    %3 = arith.divf %cst, %2 : f32
    return %3 : f32
}
```

Listing 3: Once all the necessary intrinsic functions and types are defined, regular Julia code can be used to generate MLIR.

When a developer has implemented a few intrinsic functions and types, for example in a Julia package that aims to offer a high-level interface to generate MLIR code for a particular dialect, generating MLIR code is straightforward. The top part of Listing 3 shows how a user can define a regular Julia function, `sigmoid`, and call `generate`, additionally passing in the argument type. At the bottom of Listing 3, the generated MLIR code is shown. Each function used in the definition of `sigmoid` (i.e., `/`, `exp`, ...) was implemented as an intrinsic function, similar to the intrinsic that was shown in Listing 1. During generation, these intrinsic functions will generate the operations that end up in the final MLIR code. As can be seen in the resulting MLIR code, operations of different dialects are generated. The basic arithmetic functions generate operations in the `arith` dialect, while the exponential function leads to an operation from the `math` dialect on line 4. Note also the `arith.constant` operation on line 2 that is generated to represent the literal values used in the Julia function. This is achieved by making use of Julia's type promotion system and implementing intrinsic functions for the type conversions that are implicitly called.

In MLIR, regions serve as a way to introduce structure in an otherwise flat format. They are used to group operations together and can be nested to represent control flow constructs such as loops and conditionals. Without developing extra tooling, higher-order functions in Julia can be used to support generating MLIR code with regions. To generate an MLIR operation that has a region

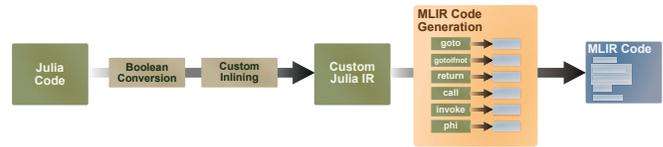

Fig. 2: Overview of the MLIR code generation process.

```
1 ─ %1 = invoke >=(_a::i64, _b::i64)::i1
 │  %2 = invoke bool_conversion_intrinsic(%1::i1)::Bool
 └─      goto #3 if not %2
2 ─      goto #4
3 ─      nothing::Nothing
4 ┄ %6 = φ (#2 => _a, #3 => _b)::i64
 └─      return %6
```

```
^bb0:
    %0 = arith.cmpi sge, %arg0, %arg1 : i64
    cf.cond_br %0, ^bb1, ^bb2
^bb1:  // pred: ^bb0
    cf.br ^bb3(%arg0 : i64)
^bb2:  // pred: ^bb0
    cf.br ^bb3(%arg1 : i64)
^bb3(%1: i64):  // 2 preds: ^bb1, ^bb2
    return %1 : i64
```

Listing 4: Julia IR and generated MLIR code for a function that returns the maximum of its two arguments. Note that basic blocks in Julia are numbered starting from 1, while MLIR block numbering is zero-based.

containing MLIR operations, a developer can simply create an intrinsic that takes the Julia function that generates this nested MLIR code and calls `generate` internally. When generating MLIR code for the outer function, MLIR generation for the inner function code will also happen.

Apart from the features already discussed, it is also possible to generate MLIR code for structured types. A function operating on a complex value `Complex{f32}`, for example, will automatically *unpack* this type until it consists of primitive types that are convertible to MLIR types, in this case, two `f32` values representing the real and imaginary part.

## IV. Implementation

In order to allow developers to implement intrinsic functions such as the one shown in Listing 1, we developed a Julia code generator tool that converts MLIR dialect specifications from Tablegen files to Julia functions that build the operation using MLIR.jl. Using these automatically generated builder functions, developers can create operations by calling functions that have docstrings and that will, to some extent, verify the validity of the arguments. This makes building MLIR operations more intuitive and much less error-prone compared to using the generic operation builder offered by the MLIR C API and MLIR.jl. Since these automatically generated builder functions are useful not only for this MLIR code generation but more generally for working with MLIR from Julia, the generator tool was upstreamed to MLIR.jl.

MLIR code generation from Julia code works by converting statements in Julia's optimised, type-inferred SSA IR. The top of Listing 4 shows the Julia SSA IR for a function that returns the largest of its two `i64`-typed arguments. Below that, the MLIR code that is generated is shown. The MLIR code generation process operates as an abstract interpreter over the Julia SSA IR. Each statement is iterated over, one by one, and interpreted in a way as to generate the relevant MLIR operations. The `>=` function in the original Julia code ends up as an invocation on line 1 in the Julia SSA IR. To interpret this statement, it is simply executed. The methods that are invoked in the Julia IR will have been implemented as intrinsic functions, and since the implementation of an intrinsic function contains the code to generate a particular MLIR operation, calling it will do just that. In the resulting MLIR code, on line 2, the invocation has given way for the `arith.cmpi` operation. Control flow statements (i.e., `goto`, `gotoifnot`, and `return`) in the Julia IR are interpreted by calling the specific builder functions (i.e., `generate_goto`, `generate_gotoifnot`, and `generate_return`). By ensuring the exact same control-flow graph (CFG) is generated in MLIR code, destinations of generated branch instructions can simply be copied over from the original Julia IR. MLIR control flow is not modeled using $\phi$-nodes like Julia SSA IR, or LLVM IR, though. Rather, branch instructions pass arguments to basic blocks to represent control-flow-dependent values. Conversion from $\phi$-nodes into this block-argument form is achieved by some additional bookkeeping in the MLIR code generation interpreter.

The Julia SSA IR shown in Listing 4 was acquired by running the Julia compiler with some altered functionality by using the AbstractInterpreter framework. First of all, the Julia programming language has no concept of user-defined types that act like a Boolean. In Julia SSA IR, the type of the condition value used in a `gotoifnot` statement cannot be anything else than a Boolean. This is a problem when generating MLIR, as different dialect might define custom types that could be used as condition values. For example, the SSA value of type `i1` that is returned by the `>=` function on line 1 in Listing 4 is used to represent an MLIR Boolean. Directly using this value as the condition to the `gotoifnot` instruction on line 3 would lead to a type inference error. To combat this, we apply a code transformation early on in the Julia compilation pipeline, before type inference has run, to insert a special call to `bool_conversion_intrinsic`. This function tricks Julia's type inference by returning a Boolean, but when this function is later encountered during MLIR code generation, the original `i1`-typed value will be used instead for generating the conditional branch operation. A second change to the Julia compilation pipeline is concerned with the inlining algorithm. By the way the MLIR code generation works, all calls to functions, except those to intrinsic functions, have to be fully inlined. Indeed, when Julia function calls

```
generate(Tuple{
        tensor{f32, 2},
        tensor{f32, 2},
        tensor{f32, 2}
    }) do C, A, B
        f = Einsum(((:i, :k), (:k, :j))=>(:i, :j))
        f(C, A, B)
    end
```

Listing 5: Julia code to generate a `linalg.generic` operation from an einsum description.

are not inlined, the MLIR code generation process has no insight in control flow that might occur within these functions. Forcing inlining ensures that the Julia SSA IR contains no hidden control flow. All basic blocks are directly visible at the top level. Intrinsic functions, on the other hand, have the opposite requirement. To avoid MLIR code generation to convert the code within an intrinsic function to MLIR code, we force these function calls to not be inlined.

## V. Case Studies

In order to evaluate the framework developed in this work, we explored case studies in three different domains.

### A. Einsum

The Einstein summation notation, einsum notation in short, is a notational convention for specifying operations on multidimensional arrays. It has uses in many branches of scientific computing ranging from tensor networks used in physics [27] to expressing new, exotic deep learning operators [28]. Matrix multiplication, for example, can be expressed as follows: $C_{ij} = A_{ik} \cdot B_{kj}$. This expression describes how a value at index $(i, j)$ in the output array $C$ can be calculated by taking the products between $A_{ik}$ and $B_{kj}$, and summing those products across all $k$.

The `linalg` dialect in MLIR is used to represent linear algebra operations. At the core of this dialect is the `linalg.generic` operation, which, in some sense, is a generalisation of the einsum notation. The `linalg.generic` operation is complex. It has a region that contains the code that specifies the computational body of the operation, and requires indexing maps and iterator types to model how the arguments are indexed. We implement an intrinsic function that can be used to create such a `linalg.generic` operation from a Julia object that stores an einsum expression. The region and other arguments are automatically generated which relieves the end user from having to specify them explicitly. Listing 5 shows how `generate` is called on a Julia function that creates a callable einsum object and calls it with three arguments of type `tensor`. This case study shows how the framework developed in this work can be used to develop simple-to-use DSLs for generating domain-specific MLIR code.

## B. Transformations and Scheduling

Transformations are fundamental to the MLIR framework. They make it possible to travel from high-level dialects to lower levels, towards executable machine code, and allow exploiting different optimisation opportunities along the way. In MLIR, transformations are predominantly implemented as *passes*, a combination of C++ and TableGen code that targets and rewrites certain operations. Applying passes is a powerful way to optimise MLIR code but often lacks the fine-grained control needed to apply transformations more locally. For example, when applied using the mlir-opt tool, the different tiling passes in MLIR will target all loops in the provided code, while a user might want to apply a tiling transformation only on a particular loop. For cases like this, a user might write more specialised passes for particular code patterns in C++. An alternative approach is to use the `transform` dialect. With this dialect, different transformations can be expressed directly in MLIR code.

The idea of the `transform` dialect is very similar to Halide [29]. In the Halide language, algorithms are implemented separately from their *schedule*, which describes how the algorithm should be executed (e.g., specifying intermediary storage, computation order, and other aspects). By making this separation, Halide allows developers to explore different schedules without changing the algorithm itself, leading to much simpler design-space exploration. Since the `transform` dialect fulfills a similar role in MLIR, it is possible to manually translate existing Halide schedules to MLIR code.

To demonstrate the ease-of-use of our MLIR code generation framework, we reimplemented a particular transformation schedule that was originally implemented as an example tutorial of porting a Halide schedule [30] to MLIR. The schedule acts on a channeled convolution operation and applies loop fusing, loop tilling and vectorisation. We succeed in matching the original schedule while writing more readable code, as the Julia code does not require explicit type annotations everywhere and we are not limited to the SSA format of MLIR. Compared to the original Halide schedule, the schedule in MLIR code, or Julia code for that matter, is still considerably less concise. However, this case study shows the potential of using a high-level language such as Julia to generate these schedules, as abstractions to quickly generate particular transformations can easily be implemented.

## C. GPU programming

From its inception, one of the main use cases for the MLIR project was to compile for hardware other than central processing units (CPUs) [3]. Most prominent is the support for targetting GPUs, as these play a central role in accelerating deep learning workloads [31]. Functions that run on GPUs are called *kernels*, and are executed by many threads in parallel. Kernels are typically programmed using general-purpose computing on graphics

```
function vadd(a, b, c)
    i = (blockIdx().x - 1) * blockDim().x + threadIdx().x
    c[i] = a[i] + b[i]
    return
end
```

Listing 6: Vector addition kernel in CUDA.jl

processing units (GPGPU) frameworks such as NVIDIA's CUDA [32], AMD's ROCm [33], or OpenCL [34]. At their core, these frameworks allow a programmer to write a kernel as a function intended to be executed by each thread mostly independently. When launching a kernel, the programmer specifies how many threads should be used, these then run the code in parallel. Originally, GPGPU frameworks were designed to be used from low-level languages such as C. However, in Julia specifically, CUDA.jl [13] allows to program kernels seamlessly from high-level Julia code to run on NVIDIA GPUs.

Listing 6 shows the definition of a simple kernel using CUDA.jl that adds two vectors together and stores the result in a third vector. The kernel uses special functions such as `threadIdx` to express thread-dependent behaviour, each thread has a different ID that is used to index the vectors. The MLIR `gpu` dialect contains similar operations to these CUDA.jl functions. After implementing intrinsic functions that generate these, it is possible to use the same exact code to generate MLIR code representing the kernel instead. Now, a call to `threadIdx` will build a `gpu.thread_id` operation, for example. The advantage over programming in CUDA.jl is that the kernel is vendor agnostic. Operations in MLIR's `gpu` dialect can be lowered to different hardware, including NVIDIA and AMD GPUs. This is similar to KernelAbstractions.jl [35], a Julia package that allows to write vendor-agnostic kernels. Currently, KernelAbstractions.jl does not offer a way to program warp-synchronous matrix-multiply accumulate (MMA) operations, however. These operations grant access to dedicated hardware in modern GPUs to execute small matrix multiplications. They have become crucial in achieving high-performance matrix multiplications. The `gpu` dialect *does* have operations to represent these, so it is trivial to write vendor-agnostic MMA kernels as well.

## VI. Conclusion

In this work, we have designed and implemented a framework that allows developers to implement high-level interfaces to MLIR dialects, in Julia. The framework builds on features of the Julia programming language such as multiple dispatch and its extensible compiler. With this framework in hand, we explored a variety of case studies. Implementing high-level interfaces for generating domain-specific MLIR code, specifying MLIR transformations in Julia, and targeting GPUs.